# Effect of severe thermo-oxidative aging on the mechanical behavior and fatigue durability of short glass fiber reinforced PA6/6.6

Florent ALEXIS[1,*], Sylvie CASTAGNET[1], Carole NADOT-MARTIN[1], Gilles ROBERT[2], Peggy HAVET[3]

[1] *Institut Pprime CNRS – ISAE-ENSMA - Université de Poitiers, UPR 3346, Département Physique et Mécanique des Matériaux - 1 avenue Clément Ader, F-86962, Futuroscope Chasseneuil, France*
[2] *Polytechnyl sas (DOMO Chemicals) – Usine de Belle Etoile, Avenue Albert Ramboz, 69190, France*
[3] *Valeo Thermal Systems – BG Material Laboratory – 8 rue Louis Lormand, 78322, Le Mesnil Saint-Denis, France*

*Email: florent.alexis@ensma.fr

## Abstract

The work deals with the fatigue lifetime estimation of Short Fiber Reinforced Thermoplastics (SFRP), with a focus on conjugated effects of thermal aging. Two materials containing 35% (V35) and 50% (V50) weight ratio of short glass fibers were aged for 500h at 200°C in air and compared to the same materials in a Dry-As-Molded (DAM) state. Monotonic and fatigue tests were performed in samples machined out of injected plates and cut along three different orientations to the injection one (0, 45, 90°) in order to capture the anisotropy of the skin-core microstructure, classical in injected SFRP.
Monotonic tensile tests evidenced the stiffening and embrittlement of the Polyamide matrix reported in the literature, nevertheless with an acuity depending on the matrix ratio and fiber orientation. Stress-controlled fatigue tests were performed at constant amplitude, frequency (10Hz), temperature (200°C) and stress ratio R=0.1. The fatigue curves of V35 are more affected by aging than those of the V50 material. The combined results from the mean strain evolutions and SEM observations confirmed that the initiation approach for fatigue lifetime estimation was still valid in aged composites. Several fatigue criteria, among the most recently reported in the literature, were evaluated from this database. The effect of aging on the cyclic evolution of the different Fatigue Indicator Parameters (FIP) involved in the criteria was analyzed preliminarily. For the present fatigue conditions, the best criteria for unaged materials (i.e cyclic creep energy-based ones) were shown to still be the best for aged materials. Finally, the ability to predict fatigue lifetime for aged composites from the identification of the fatigue criterion in the unaged state was evaluated.

## 1. Introduction

From an industrial point of view, short glass fiber-reinforced thermoplastics (SGFRP) are highly attractive materials that can replace metallic parts while respecting the intended application specifications. They represent a well-balanced solution to optimize structure lightening and the cost-efficiency ratio of manufactured parts. Thanks to the injection manufacturing process, their mechanical properties can be modulated by adjusting the fiber content. Effects of both fiber orientation [1–4] and quantity [5,6] have been extensively studied in the literature. As pointed out by several other papers, the injection parameters and the part geometrical specifications, have an important role on the final microstructure layout (skin-core-skin), and thus, in the composite properties [2,7,8]. Moreover, reinforced thermoplastics, in the same way as the matrices alone, are subjected to many external factors. One of them is the loading rate, studied in fatigue and tensile testing, through the cycle frequency [9–12] and the stress-strain imposed rates [13], respectively. The load ratio also plays a major role in the fatigue behavior of such materials, particularly for extreme values [5,14–17]. One of the consequences seen in tests performed under positive load ratio (tensile-tensile), is the so-called ratcheting effect, during which strain accumulates until complete failure of the specimens [9]. While all the above-mentioned parameters are keys in understanding the material behavior, environmental considerations are equally critical in this case.

Indeed, both short- and long-term effects occur in these materials. The former is linked to quick and reversible mechanisms such as water uptake in humid conditions and material transitions (e.g., glass transition) when heated or cooled [2,4,12,17–22]. When such conditionings are maintained for several hours, irreversible changes start to occur in the materials. Those are attributed to aging, in which, longer timeframes induce more degradation. These changes, mainly in industrial applications, are increasingly investigated since severe conditions can lead to premature failure of parts. Thus, studies are essentially focused on two aging conditions, hygrothermal [23–26] and thermo-oxidation. Both are of high concern as water and oxygen molecules are naturally found everywhere. In addition, as their name suggests, these mechanisms are driven by temperature, while composites limits are constantly pushed back due to the growing temperature requirements of applications. Hence, thermo-oxidative aging has been studied in the literature through many techniques for several aging temperatures ranging from 37 to 220 °C (for PA6 and PA6.6 matrices), and for durations up to thousands of hours in some cases. The main reported effects on the mechanical properties, stiffening and embrittlement, are mainly due to a decrease in polymer chains mobility induced by continuously self-powered oxidations reactions following the onset [27]. As reported by many authors, apart from the matrix degradation, the fiber-matrix adhesion role should not be neglected. It has been proven that proper sizing agents, containing silane [28] or polyol [29] stabilizers, have the ability to improve both thermal and mechanical aged properties. For this latter aspect, aged specimens have been compared to unaged materials for several loading conditions. Jia et al. [11] performed tensile and fatigue tests on both reinforced PA6/PA6.6 at ambient and aging temperatures (121 °C, 1000 h). While the tests at ambient temperature showed significantly diminished properties, fatigue tests performed at aging temperature were unaffected and thus, fatigue lives were preserved. Other studies focused their efforts on higher temperatures thermo-oxidative aging. Sang et al. [30] tested a commercial reinforced PA6 aged 1600 h at 90 to 180 °C, with several tests performed at 25 °C. They observed an increase in the glass transition temperature (and the crystallinity ratio) in the early stages of aging with a rapid decrease (25%) of the tensile strength. Similar results, on the mechanical aspects, were observed by Lellinger et al. [31], for a reinforced PA66 under higher aging temperatures, from 170 to 220 °C. They used an Arrhenius-type law to find an equivalent aging time-temperature dependency. Their results underlined that specimen reconditioning (i.e., humid conditions) introduced more scatter in measurements than those performed in initial conditions. These differences highlight the importance of test conditions that, in such cases, tend to limit or reinforce the aging effects. Test and aging conditions are often not the same, giving a biased view of the properties decline in real application conditions.

Hence, in this paper, we propose to study the thermo-oxidative aging effects on the fatigue behavior of newly commercial reinforced polyamides (Technyl® Red A218HPS V35 and V50). The first objective is to present the experimental fatigue database, performed at high aging temperature (200 °C, aged 500 h), from which several cyclic indicators were extracted to quantify the aging effects over the cyclic behavior and the fatigue lives. To the authors' knowledge, aging studies often addressed the effect of different matrices (e.g., PA6/PA6.6 [11,32]) or reinforcements (e.g., GF or carbon fiber [30]). Alternately, the present study focuses on the effect of fiber ratio (35 and 50% weight ratio). It was motivated by the lack of aging studies for highly filled materials (50%), especially in a fatigue context, despite the increasing interest due to their cost-performance ratio. Three different orientations with respect to the injection direction (0°, 45° and 90°) were investigated, at one temperature (200 °C) and one load ratio ($R = \sigma_{max}/\sigma_{min} = 0.1$). Specimens were dried (before aging to stay consistent with the dry environment induced by such high-temperature. Pure matrix specimens were also aged to discuss the role of the matrix in thermo-oxidative aging effects in composites. The second objective is to discuss the performances of several fatigue criteria reported in the literature, to determine the best one to estimate fatigue lives before and after aging. Recent studies compared both single-variable criteria (power-law) and multi-variable ones in a wide range of conditions including several load ratios (from -0.5 to 0.7) and different conditions (23 to 80 °C and 50 to 80% of relative humidity). Both Raphael et al. [16] and Santharam et al. [17] proposed criteria based on several fatigue indicator parameters (FIP): the hysteresis energy density, $W_h$, the cyclic energy density, $W_c$ and the cyclic mean strain rate involved in the creep energy density, $W_{cr}$. Hence, following the same strategies as these studies, criteria based on imposed stresses [1,33] and derived energies [16,17,34–39] were selected for comparison in a harsher environment.

In the following, Section 2 presents the materials and the experimental procedures (conditioning, aging, and testing) as well as the data post-processing methodology (data extraction and fatigue criteria description). Section 3 presents the tensile and fatigue results (cyclic indicators evolution and fatigue curves). Section 4 compares the capacities of the selected criteria for different identification strategies. At last, an attempt at predicting fatigue lives of the aged material with criteria identified on the fatigue database of the unaged one is presented.

## 2. Materials and experimental procedures

### 2.1. Materials

The materials studied are composites with a mixed matrix PA6/PA6.6 provided by Domo Chemicals in two glass fiber contents, 35% wt. and 50% wt. respectively commercialized under the names Technyl® Red A218HPSV35 and Technyl® Red A218HPSV50. They were prepared by loading short glass fiber reinforced thermoplastic pellets in an injection molding machine, forming 360 x 100 x 3 mm³ rectangular plates with the longest dimension being the injection direction. Plates were injected in dry-as-molded (DAM) conditions and dogbone-shaped samples (Fig. 1b) were extracted in three positions on each plate by Domo Chemicals following three orientations compared to the injection direction ($\theta°$): 0°, 45°, and 90° (Fig. 1a). The through-thickness evolution of the first component of the second-order orientation tensors ($a_{xx}$, along the *x* direction aligned with the injection one) is given for both composites (Fig. 1c). This will be used as a reference in the result sections. Additionally, plates without fibers (0% wt.) were also provided by Domo Chemicals and machined directly in the laboratory facilities following the ISO 527-2 norm, which is a usual design for pure polymer specimens (Fig. 1b).

### 2.2. Conditioning: initial state and aging

Thermoplastics materials are known to be highly sensitive to their environment. Temperature and humid conditions can both have short-term (ductile-brittle transition) and long-term (aging) effects on the material's behavior, especially on components close to the engine. Hence, to ensure tests reproducibility and homogenous conditions, all samples were dried under vacuum for 24 h at 80 °C and sealed before testing. These samples are referred to as "unaged" in the following sections. After drying, some samples were aged in an oven at 200 °C under air (natural composition, for normal rate oxidation) for 500 h and sealed again shortly after.

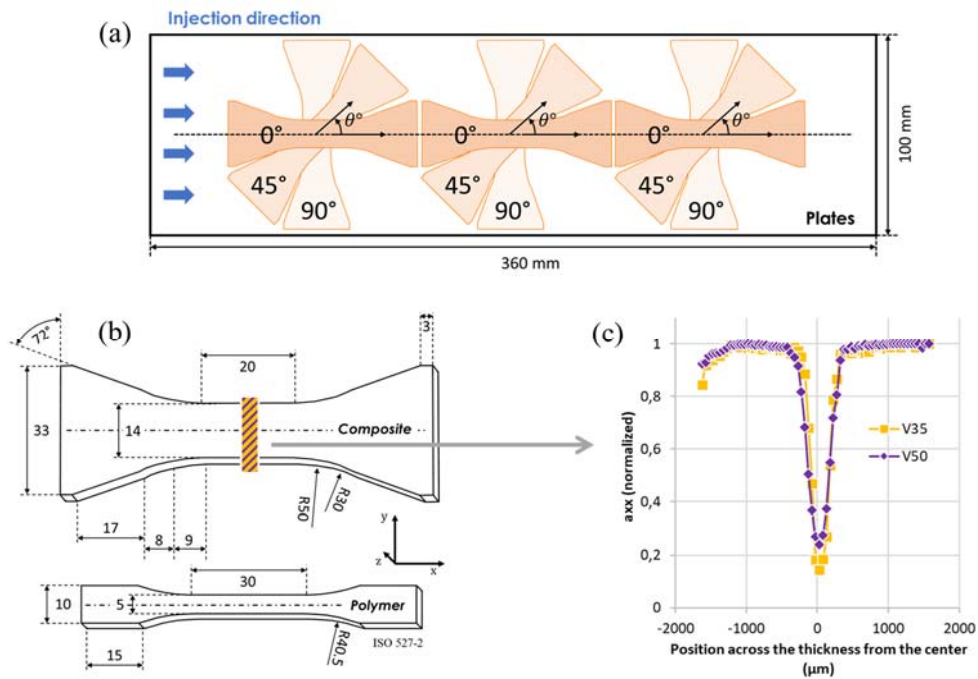

*Figure 1 Specimens' specifications: (a) orientations and positions of composite samples machined out of injected plates comparatively to the injection direction; (b) dimensions of the machined specimens (mm) and (c) the evolution of the averaged $a_{xx}$ component (x direction aligned with the injection one) of the second-order orientation tensor measured in the plates' thickness obtained from micro-tomography for both materials (35 and 50% wt.)*

### 2.3. Experimental tests and techniques

Mechanical tests were performed on a servo-hydraulic Intron® 8802 fatigue testing machine (5 kN load cell) equipped with a climatic chamber heated at aging temperature (200°C) in air. Temperature was stabilized within one hour of heating (13 °C/min ramp) and tests were performed in series without waiting for the complete cooling of the machine's loading train (measured by a thermocouple). The quasi-static tensile tests were controlled by crosshead displacement equivalent to a nominal strain rate of $10^{-3}$ s$^{-1}$. The fatigue tests were conducted exclusively on composites specimens, under load control at a fixed loading ratio ($R = \sigma_{max}/\sigma_{min} = 0.1$) in a sinusoidal waveform, a constant amplitude, and a number of cycles to failure ($N_f$) ranging from $10^3$ to $10^6$. Before each test, the initial cross-section dimensions of the sample were measured (averaged value over multiple measurements) and taken into account to calculate the stress level, especially because dimensional changes were observed after aging. On the other hand, the evolutions of cross-sections during testing (thermally induced expansion) were not accounted for. The strain evolution was monitored by a digital image correlation (DIC) setup positioned outside of the climatic chamber and tracking four painted markers on the sample surface. The chosen fatigue frequency was initially set at 10 Hz. It is assumed that self-heating effect did not significantly affect the results given the test temperature (200 °C). Setup constraints linked to the camera refreshing rate limited the data output at this frequency. Thus, slower blocks at 1 Hz were inserted in-between 10 Hz blocks, every 200 cycles at the beginning and every 1000 cycles after $2 \cdot 10^4$ cycles, in order to acquire sufficiently well-described loops with at least 35 data points. A time gate was needed to ensure stabilization after blocks transitions due to the delayed material response (short-term relaxation linked to the viscoelastic behavior). Preliminary tests aiming at setting the right set of PID controller parameters (proportional, integral, and derivative) showed that 6 cycles at 1 Hz were sufficient to achieve the imposed stress levels. The indicators deduced from hysteresis loop or the mean strain increment, reported in the Results section, were evaluated at the 6$^{th}$ cycle. An example of monitored data for one fatigue test is presented in Fig. 2a. The fatigue loops, $\sigma - \varepsilon$, are drawn with markers showing measured data points (for the cycles at 1 Hz) and the mean strain ($\varepsilon_m$) evolution measured with 10 Hz data points is plotted as a function of the number of cycles ($N$) using a secondary Y-axis (right side). Transitions 1Hz-10Hz (pointed by an arrow) induce peaks in strain values that quickly (few cycles) decrease to reach stabilization. As shown in Fig. 2b, the 1 Hz cycles represent less than 3% of the total number of cycles to failure ($N_f$) for short tests and this fraction decreases all the more the test duration is long. The impact of such 1 Hz acquisition blocks on fatigue lifetime will be detailed in the Results section (section 3). The data were monitored until the complete failure of the specimens. For confidentiality reasons, all stress-related values (ex. nominal stress and imposed maximum stresses for the S-N curves) were normalized by a common arbitrary factor whereas all the other quantities (e.g., nominal strain, and number of cycles) were unchanged. In this way, comparative analysis between materials (fiber concentration, unaged/aged) is still valid. In addition, for the sake of representation here, values in the aged state were divided by the values in the unaged one (i.e., the reference) to better compare the aging influence on mechanical quantities. They are referred to as "ratios" in the following sections.

A scanning electron microscope (SEM) from Tescan® (VEGA3), was used to observe failure surfaces of aged specimens.

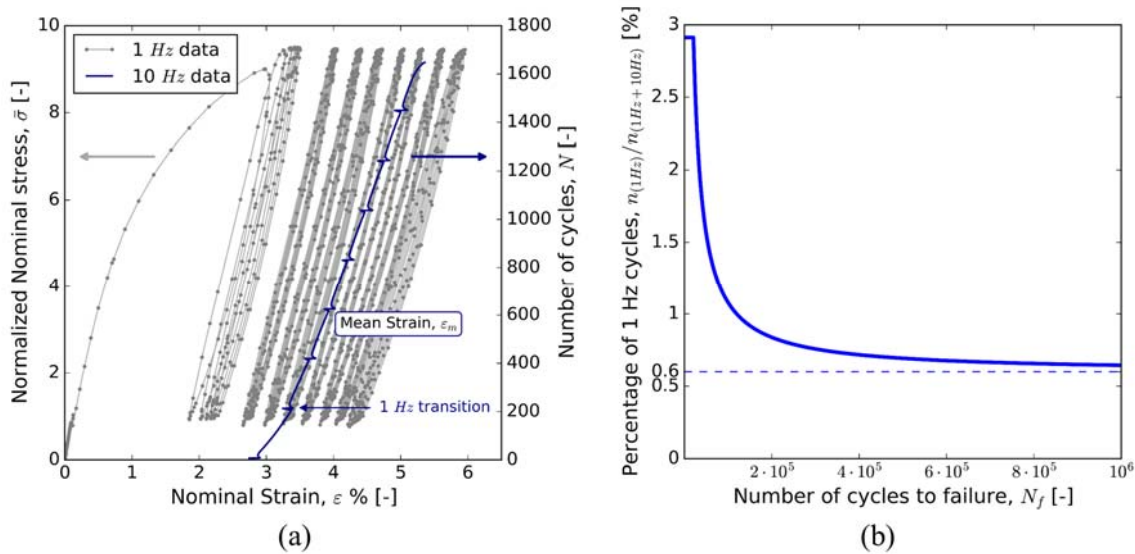

*Figure 2 (a) Example of monitored data: fatigue loops (for the blocks at 1Hz) and evolution of the mean strain measured at 10 Hz and (b) evolution of 1 Hz cycles' proportion as a function of the number of cycles to failure, $N_f$.*

### 2.4. Cyclic indicators

As mentioned in the introduction, several fatigue criteria are based on mechanical parameters, called Fatigue Indicator Parameters (FIP), which can be obtained from the analysis of hysteresis loops in the case of uniaxial fatigue loading. These quantities are usually taken at mid-life, considered as a stationary state, and will be noted with a * symbol in the following sections. The FIP will be plotted against the experimental number of cycles to failure, $N_f$. Those used in this study are described in Fig. 3.

*Apparent modulus.* The apparent modulus ($E_{app}$) or secant modulus is the ratio between the stress amplitude ($\Delta\sigma$) and the strain amplitude ($\Delta\varepsilon$) where $\Delta$, the amplitude, is defined as $\Delta\aleph = \aleph_{max} - \aleph_{min}$ (Fig. 3a).

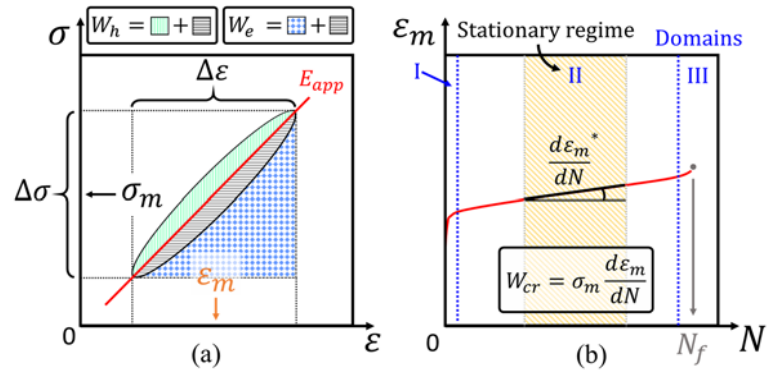

*Figure 3 Mechanical quantities employed in the present study for fatigue criteria computation: (a) determination from the hysteresis loop and (b) from the axial mean strain evolution until failure ($N_f$).*

*Cyclic mean strain rate.* The cyclic mean strain rate corresponds to the slope of the mean strain ($\varepsilon_m = (\varepsilon_{max} + \varepsilon_{min})/2$) against the number of cycles ($N$). It can be written as $\dot{\varepsilon}_m = d\varepsilon_m/dN$. For SGFRP under DAM and humid conditions, the mean strain value increases in a non-linear way with fatigue loading ($R \geq 0$) [20]. As illustrated in Fig. 3b, the slope decreases rapidly after the first 100 cycles (regime I), reaching a stationary regime (II) followed by a quick increase near the final rupture (regime III; 90% of $N_f$). In this work, the strain used in the computation is the nominal axial strain measured in the loading direction ($x$).

*Creep energy density.* The creep energy density ($W_{cr}$) is the product of the cyclic mean strain rate by the cyclic mean stress ($\sigma_m$) (Fig. 3b).

*Hysteresis energy density.* The hysteresis energy density ($W_h$) corresponds to the cycle area (Fig. 3a), it is linked to the dissipated and stored energy density of the fatigue-loaded material. Experimentally, the trapeze method is used for computation and this quantity can be written as $W_h = \int_{cycle} \sigma d\varepsilon$.

*Elastic strain energy density.* The elastic strain energy density ($W_e$) is equivalent to the area under the apparent modulus (Fig. 3a). It can be written as $W_e = \frac{1}{2}\Delta\sigma^+\Delta\varepsilon$. It is worth noting that this quantity can only be computed for positive stress and strain values, otherwise, one must use the cyclic energy density ($W_c = \frac{1}{2}\Delta\sigma^-\Delta\varepsilon$) for negative load ratios, where at least one of the stress components is negative.

These FIP are involved in different types of criteria listed in Table 1; they obey either simple power-law forms or more complex equations with more than two parameters.

| *Criterion* | *Equation* | $n_p$ |
|---|---|---|
| *Power-law, $\sigma_{max}$* | Power-laws under the form: | |
| *Power-law norm., $\sigma_{max}/\sigma_{UTS}$* | | |
| *Elastic strain energy, $W_e^*$* | $FIP = \beta \cdot N_f^\alpha$ | 2 |
| *Hysteresis energy, $W_h^*$* | with **FIP** being the Fatigue Indicator Parameter | |
| *Creep energy, $W_{cr}^*$* | | |
| *Mixed criterion, IRvar1 [16]* | $N_f = \min\left[A \cdot \left(\frac{d\varepsilon_m^*}{dN}\right)^n ; B \cdot W_h^{*m}\right]$ | 4 |
| *Mixed criterion, IRvar2 [16]* | $N_f = (1-w) \cdot \left(\frac{d\varepsilon_m^*}{dN}\right)^n + w \cdot W_h^{*m}$ | 3 |
| *Mixed criterion, IDAFIP1 [17]* | $N_f^{-1} = \left(\frac{W_{cr}^*}{A}\right)^\alpha + \left(\frac{W_h^*}{B}\right)^\beta$ | 4 |
| *Mixed criterion, IDAFIP2 [17]* | $N_f^{-1} = \left(\frac{W_{cr}^*}{A}\right)^\alpha + \left(\frac{W_e^*}{B}\right)^\beta$ | 4 |

*Table 1 Tested fatigue criteria: equations and number of parameters, np.*

### 3. Experimental results and discussion

#### 3.1. Tensile results

In this part, tensile responses of matrix (0% wt.) and composites (35, 50% wt.) specimens in both conditionings (unaged, aged) are presented at aging temperature (200 °C). Matrix results are only used for comparison. The results are reported in Fig. 4. Only one curve per condition is represented for clarity reasons although several tests were performed for reproducibility concerns. It is to be noted that the color code used for each orientation will be the same throughout the paper. Since the unaged matrix is highly deformable, tests were interrupted at 100% of deformation, long before complete failure and the apparition of necking commonly observed for such materials. In order to keep similar X-axis scaling until 20% of deformation for the 3 subfigures, the last increments of deformation for the matrix are plotted using a broken X-axis (Fig. 4a).

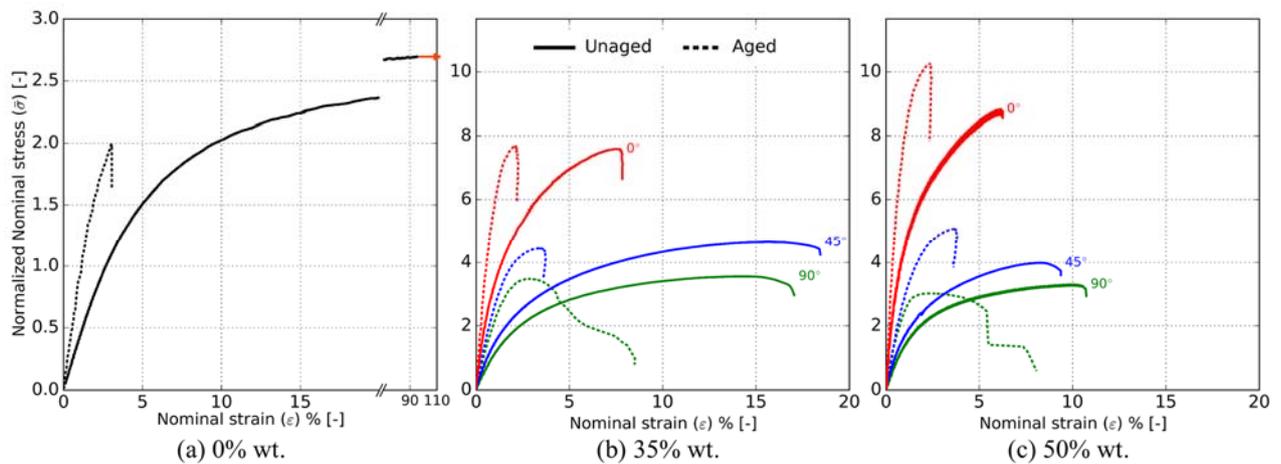

*Figure 4 Tensile tests results performed at 200 °C of unaged (plain lines) and aged (dotted lines) (a) 0%, (b) 35% and (c) 50% wt. fiber specimens.*

Mechanical properties (tensile modulus, maximum stress and strain to failure) for a given material, orientation, and conditioning, were issued from arithmetic averages over all the tests performed for this material in the same conditions. The resulting values in the aged state were then divided by their counterparts in the unaged state (acting as a reference). Fig. 5 represents the values of the ratios thus obtained for the tensile modulus, the maximum stress and the strain at failure for the matrix and each composite and orientation.

Consistently with reported results in the literature [31,32] regarding the impact of thermo-oxidative aging, the matrix as well as the composites whatever the orientation, appear more rigid and less deformable after aging. The tensile modulus is almost doubled after aging for every tested material, with the 90° orientation value being closer to the matrix one as shown by the evolution of the modulus ratio. It is worth mentioning that, for the unaged materials, the 90° specimens of the 35% wt. composite appear more rigid than the corresponding specimens of the other material. This is counterintuitive as less reinforcements would indicate a softer material. This difference is due to the microstructure, as in the case of the former material, more fibers, in the core layer, are aligned with the tensile direction as shown by Fig. 1c. Even though the fiber concentrations are quite different, this leads to stiffer specimens in this case (about 18% in average). After aging, with the decrease in angle $\theta°$ (from 90° to 0°), the composite rigidity decreases in the same way for both fiber contents. This is probably due to the ratio of load transferred between the matrix and fibers, more rigid when most of the fiber are aligned with the loading direction and vice-versa, which ultimately depends on the specimen's orientation. In this case, the orientation effect seems to take over the fiber content one. For tests at room temperature, it is shown in the literature [31] that the maximum stress and the stress at failure significantly decrease after aging due to much earlier fracture (embrittlement induced by aging). For tensile tests performed at 200 °C, only pure matrix specimens exhibit such a behavior. In the unaged state, it is expected that ductility should decrease with the increase of fiber content (and specimens' orientation) leading to a predetermined ranking order between the tested materials (0%>35%>50% wt. in deformability). After aging, the complete opposite is observed. The matrix specimens are no longer capable of high deformation and necking no longer occurs before the final failure. For the 35% wt. material, the aging decreases the orientation effect on the failure strain, while the effect of orientation remains fully expressed for the aged 50% wt. composite. At last, aging disturbs the effects of orientation on the maximum stress values and no real tendency can be extracted. The maximum stress of the 35% wt. composite is roughly the same in both states whatever the orientation, while only the 90° 50% wt. specimens exhibit such a characteristic. For the other orientations (0° and 45°) maximum stress values are higher after aging as if the material was less degraded. In other words, for a fixed stress level, the departure from the monotonic stress at failure is the same in the 35% wt. specimens in both conditionings (pristine and aged) whereas it is not the case for the 0° and 45° specimens of the 50% wt. composite where the stress at failure is raised after aging. As strain and stress are both linked, higher ratios of the strain at failure for the 0° and 45° specimens of the 50% wt. composite could explain the increase of maximum stress values. Moreover, the failure in the case of 90° specimens appear to be less sudden than for other orientations, with an ever-decreasing stress value after the maximum stress. These differences could be linked to surface micro-cracks induced by aging or to the propagation of internal cracks through the aged brittle matrix phase. These results are very important as they will serve for further discussion of the fatigue curves and cyclic indicators evolutions presented in the following sections. It is worth noting that, since some specimens (0° and 45°, 50% wt. composite) appear more resilient to aging (or at least less degraded), the S-N curves which rely on the imposed stress should be heavily impacted by such an increase of the maximum stress.

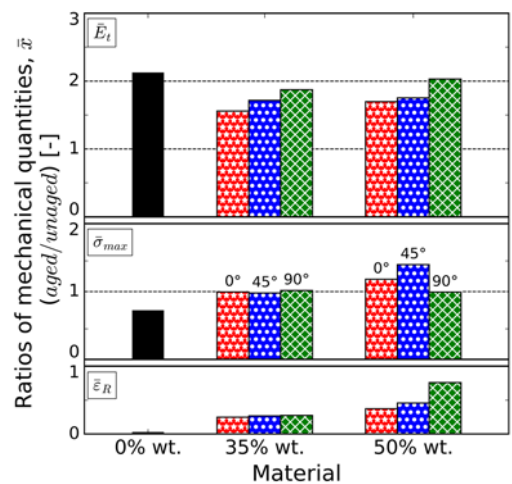

*Figure 5 : Evolution of tensile properties: tensile modulus, maximum stress and strain at failure for the matrix, both composites and the 3 orientations. Each quantity after aging is divided by its value in the unaged state.*

### 3.2. Fatigue results

In this part, the fatigue behavior of unaged and aged composite specimens is investigated. The database was constructed only at aging temperature (200 °C) to mimic part of service conditions in industrial components, from accelerated tests.

*3.2.1. S-N curves*

S-N curves are usually plotted to compare the fatigue life ($N_f$) against the imposed stress ($\sigma$). The used stress component varies depending on the database. In this study, since only one load ratio is studied, the maximum stress is employed. The values of imposed maximum stress ($\bar{\sigma}_{max}$) are normalized by the same arbitrary factor whatever the orientation and conditioning.

First of all, the influence of 1 Hz acquisition blocks on the fatigue life had to be evaluated. Little to no effect was expected on the fatigue lives as 1 Hz cycles represent only a little fraction of the total number of cycles (as shown in Fig. 2b). In order to verify this assumption, fatigue tests at 10 Hz were conducted, for two stress levels per orientation chosen to obtain fatigue lives near $10^4$ and $10^5$ cycles. The obtained fatigue lives (with 2 tests for each stress level) for the 50% wt. unaged specimens are reported (in blue) in Fig. 6 representing the S-N curves at alternated 1-10 Hz frequency. The values at 10Hz appear to be within the usual dispersion interval known for this type of material. Even if only 4 points is low for a given orientation, the set of data (12 points) seems enough to show consistent results across all tested orientations.

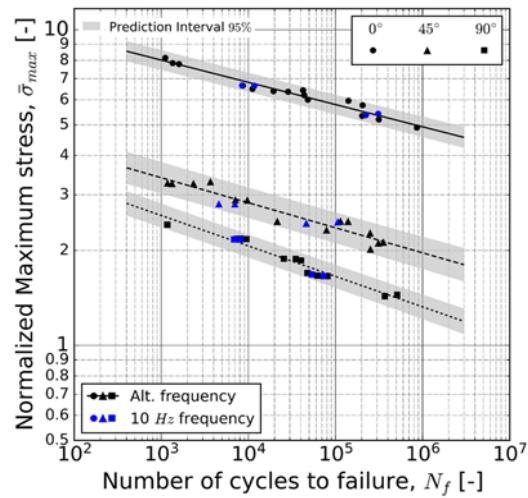

*Figure 6 Normalized fitted S-N curves at alternated 1-10Hz and data points at 10 Hz, 50% wt. fiber composite and 3 orientations.*

The S-N curves of the 35% and 50% wt. composites are reported in Fig. 7. The fatigue data are fitted using logarithmic regressions and both 95% confidence and prediction intervals are drawn only for the unaged composites for clarity reasons. Similar dispersions are found for all data sets. The width ratios of every confidence interval, named $T_N$ ($= [l]N_{f,10^6}/[r]N_{f,10^6}$, with $l$ and $r$ being respectively the left and right bounds), were computed and are given at the extremity of each curve. They indicate that the dispersive nature of fatigue data points is unaffected by the considered aging conditions. The identified power-law parameters, $\alpha$ (slope) and $\beta$ (intercept) (see equation in Table 1) for every state and orientation, are given in the tables in the corner of each figure. The stiffer character of the less fiber-charged material for the 45° and 90° orientations is fully expressed, consistently with the quasi-static results (subsection 3.1). It can be observed that thermo-oxidation tends to decrease the slope and increase the intercept which, at a fixed stress level, essentially leads to shorter fatigue life. The amplitude of variations gradually increases with the angle $\theta°$, with the worst impact of aging on the 90° specimens regardless of the fiber content.

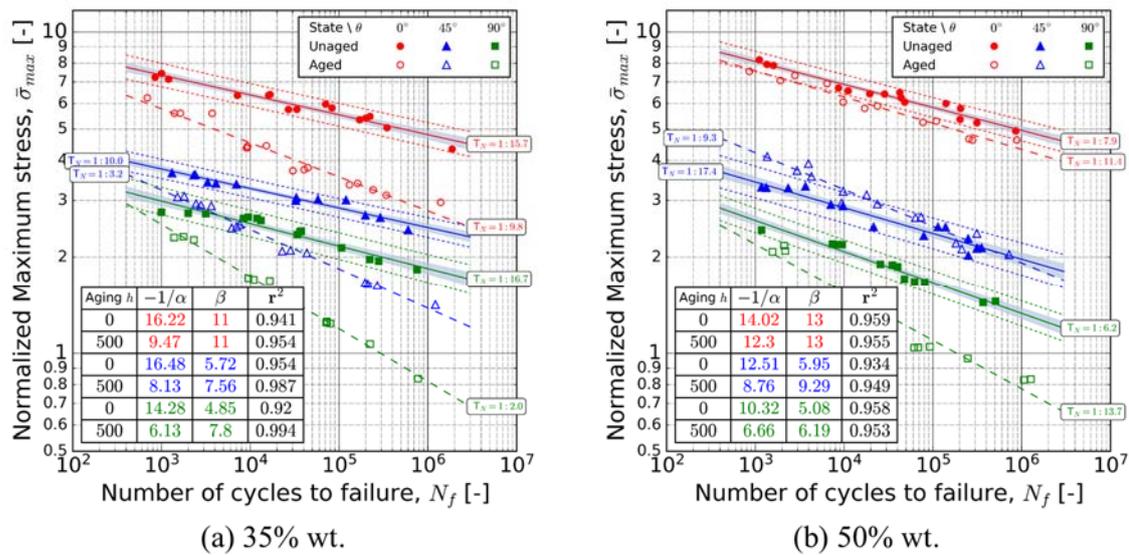

*Figure 7 Influence of fiber content on the fatigue behavior through normalized S-N curves: (a) 35% wt. fiber composite; (b) 50% wt. fiber composite.*

To better highlight the differences in the two composites, the ratios of the values in the aged state by those in the unaged one (Fig. 8). As pointed out by this figure, the only outlier is the value of $\beta$ for the 45° 50% wt. material, for which a peak is observed due to the fact that the S-N curves for both unaged and aged states are close but with different slopes. Tensile results showed a worsened aging impact for the 35% wt. specimens, with disparities in the maximum stress values and the strains at failure that were strictly inferior to those observed for the higher fiber ratio. As suggested in the previous part, the aging-increased maximum stress observed for 0° and 45° 50% wt. composites may be responsible for the better fatigue lives after aging in comparison with the 35% wt. material. The comparison between Fig. 7a and b for 0° orientation shows the higher impact of aging for the 35% wt. composite. Taking $N_f = 10^6$ cycles as a reference value for a given stress level on the S-N curves related to the unaged state (Fig. 7), drops to $5.8 \cdot 10^3$ (5.8 thousandths of the unaged fatigue life) and $2 \cdot 10^5$ (1 fourth) cycles are seen for the 35% and 50% wt. specimens, respectively. Aging is much more detrimental in the 35% wt. material's case which can be directly illustrated through the slope evolutions (aged/unaged) of both composites for the 0° orientation (Fig. 8).

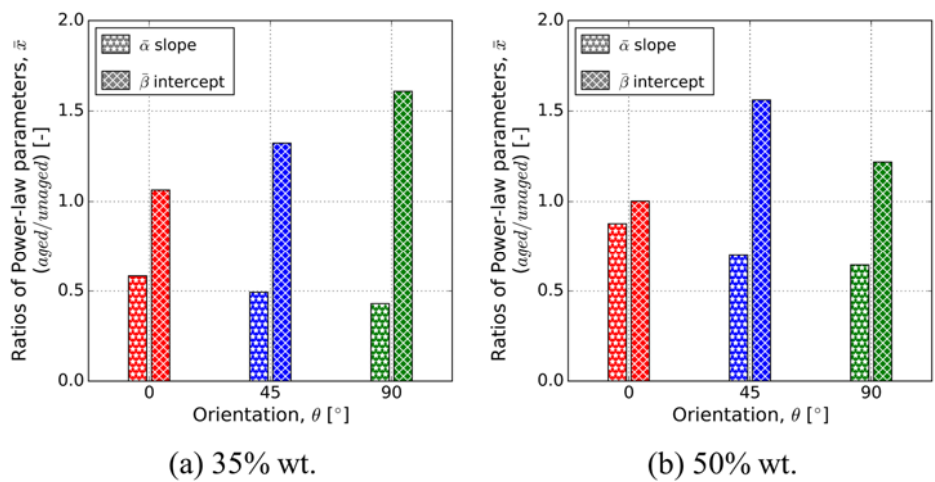

*Figure 8 Evolution of slope and intercept power-law parameters as a function of orientation: (a) 35% wt. fiber composite; (b) 50% wt. fiber composite. Each quantity after aging is divided by its value in the unaged state.*

To the author's knowledge, such disparities in aging impact due to the matrix to fiber weight ratio combined with the process-induced microstructure were not reported in the literature. This is undoubtedly related to the fact that very few studies investigated the aging effect in the very same conditions as the initial conditioning as mentioned in the introduction section.

### 3.2.2. Aging impact on cyclic indicators

From the above comparisons, the thermo-oxidative aging impact seems to depend heavily on the microstructure (orientation) and on the fiber content, with some similarities between the tensile and fatigue responses. In a fatigue context, where multiple criteria rely on the consideration of a stationary regime (linearly increasing strain rate), aging impact on cyclic indicators must be analyzed thoroughly. Thus, in Fig. 9, monitored data such as hysteresis loops (measured at mid-life), and mean strain evolutions are presented for the 0° (top subfigures) and 90° (bottom subfigures) 50% wt. composite to illustrate the extreme cases. The nominal stress is normalized by the same factor for every loop (also the same one used for the maximum stress figures). Similar trends were observed for the other orientation as well as for the other composite.

In each figure, filled and unfilled markers refer to the unaged and aged materials, respectively. A logarithmic gradient color bar is plotted to specify the fatigue life of each colored data set.

*Hysteresis loops.* Fig. 9a shows the stress-strain hysteresis loops measured at mid-life. Hysteresis loops have an asymmetric shape for both conditionings. The loop area, and so the corresponding hysteresis energy density, decreases with the imposed stress amplitudes (Fig. 9a). Hence, loops of the longest tests (red color) have near-zero energy when compared to shorter tests (blue color; $< 10^4$ cycles). Similarly, the apparent modulus and the elastic strain energy density are also decreasing. Following the results observed on the S-N curves (Fig. 7), at 0°, relatively close stress amplitudes are found between both conditionings for tests of equivalent duration. On the other hand, with a greater gap visible for the 90° orientation, the resulting stress values tend to be positioned at lower levels for the same fatigue lives. Comparatively, the strain amplitudes are heavily decreased whatever the orientation. These differences are consistent with the evolutions previously observed in the S-N curves. Essentially, the closer the S-N curves in both states, the closer the stress amplitudes. Thus, the aging impact can be distinguished in two effects. The first aging effect, in priority order, is the reduction of strain by embrittlement which impacts directly the derived energies. At a more advanced stage, another change occurs, the stress amplitudes required to achieve a particular lifetime tend to decrease, inevitably leading to an even greater decrease of loop-derived energies. For the 35% wt. material, for which larger gaps between the S-N curves of the unaged and aged states were found independently of the orientation, the tendencies previously described for the 50% wt. composite are exacerbated. This is consistent with the stiffening and embrittlement observed on tensile curves (Fig. 4).

*Mean strain evolution.* Fig. 9b presents the mean strain evolution as a function of the number of cycles to failure. The X-axis bounds were chosen to show the majority of the dataset range except the very end of the longest tests ($> 3 \cdot 10^5$ cycles). An inserted graph focuses on the shortest tests ($< 2 \cdot 10^4$ cycles) for clarity purposes. Despite a clear reduction of strain amplitudes, linked again to an increased rigidity after aging, the three strain rate regimes (I→III, see Fig. 3) still can be distinguished after aging. Moreover, and very importantly for the following, the strain kinetics of unaged and aged specimens exhibiting similar fatigue lives seem close to each other, relatively to the respective strain amplitudes, in the stationary regime II.

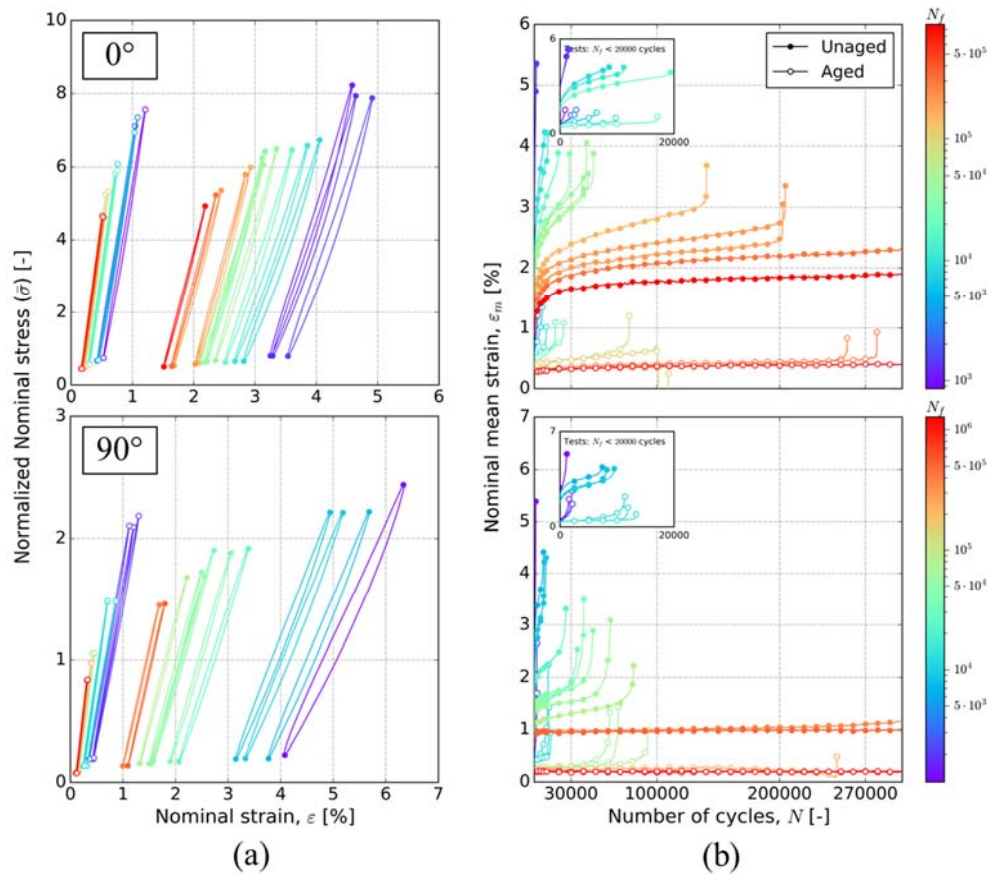

*Figure 9 Aging impact on cyclic indicators on the 0° and 90° 50% wt. fiber specimens: (a) mid-life hysteresis loops and (b) mean strain evolution during fatigue tests.*

*FIP evolution.* The values of a given FIP at mid-life ($W_e^*$, $W_h^*$, and $d\varepsilon_m^*/dN$) for the three orientations were plotted as functions of the number of cycles to failure, as illustrated in Fig. 10a for the 35% wt. composite. Logarithmic regressions were computed over the 3 orientations for each material and conditioning. Then, values of the logarithmic regression in the aged state were divided by the ones in the unaged state for fatigue lives ranging from $10^3$ to $10^6$ cycles. Values of these ratios are represented in Fig. 10b as a function of the number of cycles to failure for the three FIP and for both composites. The taller the bar, the less difference between aged and unaged values.

This graph reveals a lesser impact of aging on the hysteresis energy density, $W_h^*$, for both composites. Similarly to the above tests results (quasi-static and fatigue), a larger gap between the unaged and aged states can be observed for the 35% wt. composite. Moreover, the influence of aging on the three FIP increase with fatigue lives. In the case of the 50% wt. composite, the influence of aging on the cyclic mean strain rate is relatively stable, and lower than on the elastic strain energy density. Meanwhile, the 35% wt. material shows continuously decreasing values, with the cyclic mean strain rate being the most impacted indicator. These results show that aging has a heavily different impact on the indicators.

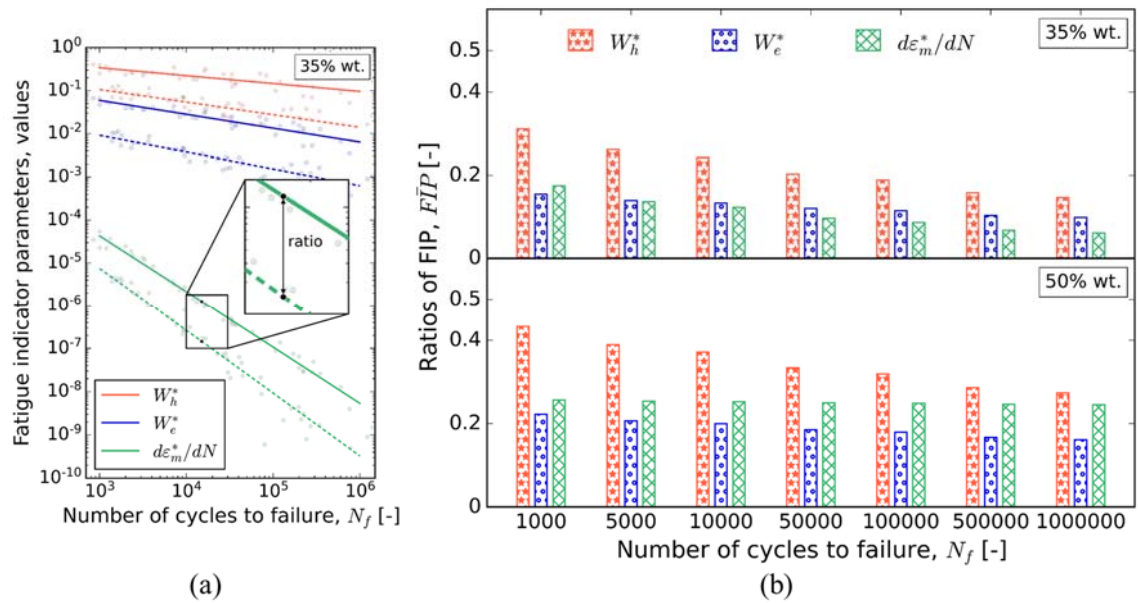

*Figure 10: (a) Hysteresis energy, elastic strain energy, and mean strain rate (at mid-life) as functions of the number of cycles to failure with logarithmic regressions over the 3 orientations for each conditioning (35% wt. fiber composite). Data points are represented in light grey color (without distinction of the orientation) and the regressions for the unaged and aged states are drawn using plain and dotted lines, respectively. (b) Evolution of FIP as a function of the number of cycles to failure (35% and 50% wt. fiber composites). For a given fatigue life, each FIP value in the aged state is divided by its value in the unaged state, both values being issued from the previous regressions.*

Finally, apart from the cyclic indicators, and especially since thermo-oxidative aging is a non-homogeneous mechanism (oxidized layer with formation of surface cracks), special attention was paid to the failure surfaces during the database construction. Whatever the conditioning, these surfaces were almost exclusively localized in specific regions of the specimens as shown in Fig. 11 meaning the preponderant role of the specimens' geometry (Fig. 1b) on the failure location. The surface aspects were nevertheless dependent on the conditioning, being rougher and more tortuous after aging. Additionally, an extension of such aging conditions for periods up to 2000h showed the formation of critical cracks around the core layer preventing any fatigue testing in repeatable conditions. In Fig.11, two images of fatigue-loaded aged specimens at two different fatigue lives are presented to show the crack initiation origin and the packed spherical-cavities that can only be observed after aging. These cavities were also visible on the monotonic tensile specimens loaded at aging temperature. Further investigations on the aged specimens at ambient temperature were made to confirm that they are induced by temperature and rather different from the cavities that may appear during the "cavitation" damage mechanism described in the literature for fatigue loading [40]. For this orientation, the crack initiation seems to be localized in the core region, with a clear transition between two domains with different aspects (rough to smooth), which is consistent from the literature results regarding the location [33,40] and the ratio of microductile (rough) to microbrittle (smooth) area when the stress (and so the fatigue lifetime) decreases [33]. From the combined results regarding the evolution of cyclic indicators (previous section) and the post-mortem analysis, it is safe to assume that aged specimens exhibit the same damage mechanisms as unaged ones. Indeed, the formation of the mesocrack, visible to the naked eye and responsible for the final rupture, has been shown to occur near the end of the fatigue lifetime when a deviation of the maximum strain is observed (start of regime III, Fig. 3) [40]. As demonstrated by the authors, it originates from the coalescence of several damage markers (fiber ends, microcracks…) and the creation of bridges between those. Since the herein discussed aging conditions have no impact on the mean strain regimes repartition and the crack origin, the initiation approach, where the majority of the fatigue life is on homogenous damage before initiating a crack, can be used as well for fatigue lifetime estimation on the aged database.

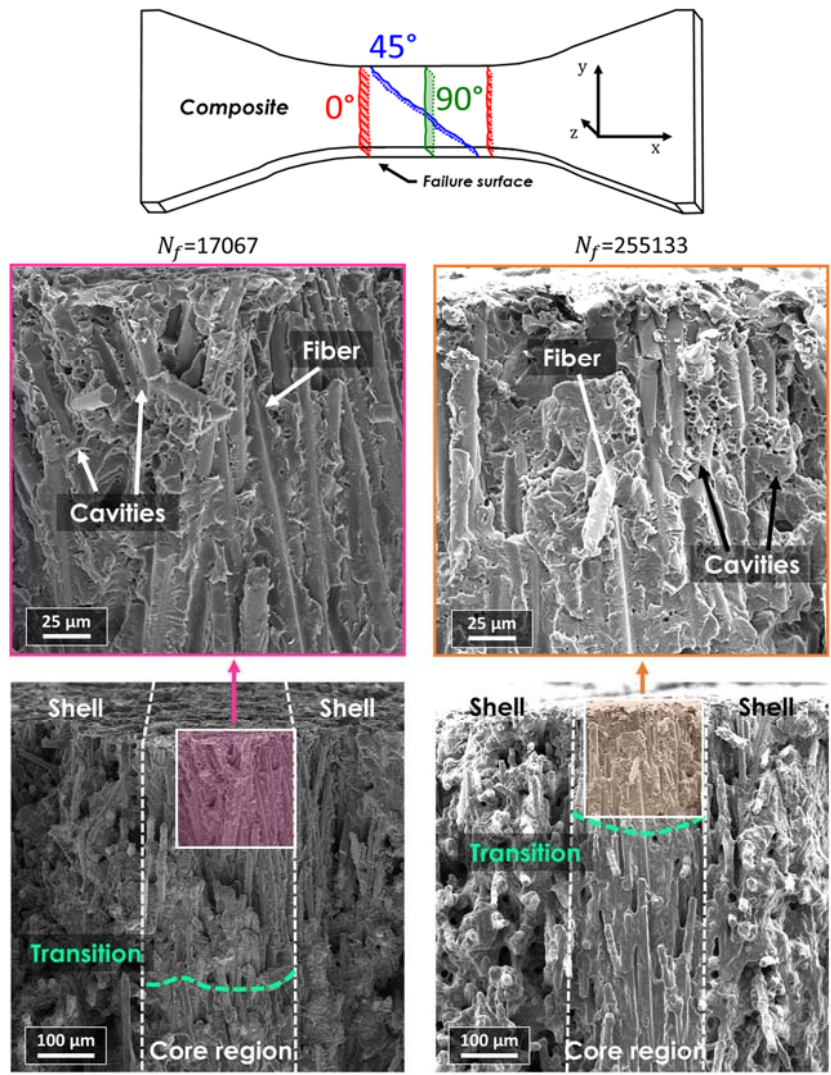

*Figure 11 Failure paths dependent on the specimen orientation obtained for both unaged and aged composites (top subfigure) and SEM fractography of the failure surfaces (with a close-up areas near the edges, i.e., along the Y direction) of fatigue-loaded aged specimens (0°, 50% wt. fiber content) for two numbers of cycles to failure.*

## 4. Fatigue criteria

The second aim of the work was to investigate the thermo-oxidative aging influence on the performances of a wide list of criteria (see Table 1). The first concern deals with the ability of each criterion to unify the fatigue lifetimes for each conditioning separately; the criteria parameter identification is performed separately on each material state (subsection 4.1). The second step, attempts at predicting fatigue lives of aged composites from the knowledge of fatigue data in the unaged state (subsection 4.2). The two composites (35% and 50% wt. composites) will be compared throughout the section.

At first, the fatigue curves, using the mid-life cyclic creep energy ($W_{cr}^*$), are reported in Fig. 12 for both materials. The results are presented in the same way as the previous S-N curves (Fig. 7), with a set of parameters identified for each conditioning and orientation. It can be noticed, through the $r$ squared values (close to 0.99 in most cases), that the overall data point dispersion is significantly reduced when compared to the stress-based criterion. The curves are also closely packed, contrasting with the previous results, which were dependent on the specimens' orientation. In the family of power-law type criteria, $\bar{\sigma}_{max}$ and $W_{cr}^*$, represent both extreme cases, performance-wise, when looking at our available set of FIP. Even with $W_{cr}^*$ the best FIP reported in the literature for positive load ratios [9], results for the two composites are distinct. Indeed, while the 35% wt. composite has two separate sets of curves depending on the material conditioning (Fig. 12a), the 50% wt. composite possesses overlapped data (Fig. 12b). This difference comes from the advanced degradation observed for the less fiber-charged composite for both tensile and fatigue tests (Sections 3.1 & 3.2). Moreover, similarly to the S-N curves (Fig. 7) a decrease in the slope is observed after aging, but no clear tendency can here be extracted from the intercepts.

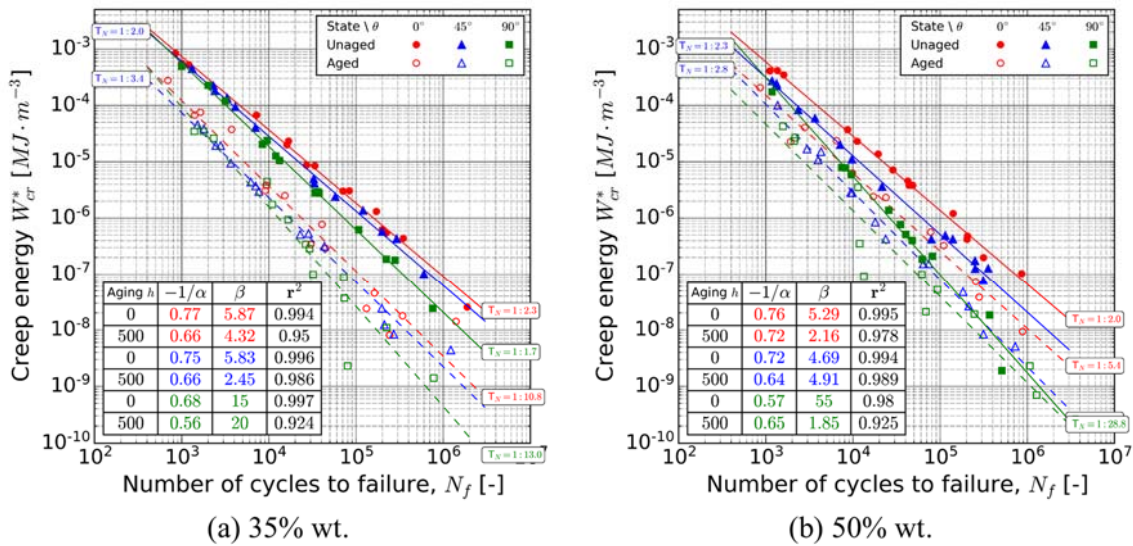

*Figure 12 Influence of fiber content on the fatigue curves using the mid-life cyclic creep energy, $W_{cr}^*$: (a) 35% wt. fiber composite; (b) 50% wt. fiber composite.*

### 4.1 Ability of criteria to unify fatigue lifetimes for each conditioning separately

The first strategy consists in using a unique set of parameters identified for each material state. The three orientations were gathered into two databases under the respective names, *Unaged* and *Aged*. The number of cycles to failure ($N_{f,calc}$) was then computed for each conditioning with the parameters corresponding to each conditioning and plotted against the experimental fatigue life ($N_{f,exp}$) in an equivalent graph. This was done for both composites. An example for the 35% wt. composite is given in Fig. 13 for the mid-life cyclic creep energy ($W_{cr}^*$) based criterion. In Fig. 13a, the cyclic creep energy $W_{cr}^*$, is plotted as a function of $N_{f,exp}$ and the identified parameters are given in the inset table. Fig. 13b represents $N_{f,calc}$ vs. $N_{f,exp}$ with both conservative and non-conservative domains and the percentages of data points inside each scatter band (factor 2, 3, and 5). For each conditioning, *Unaged* or *Aged*, this criterion is good at unifying the fatigue data obtained at 200 °C for the 3 orientations. The $r$ squared values and the $T_N$ ratios (Fig. 13a) indicate a relatively small data dispersion. Almost all data points (97.7%) are included in the larger band (factor 5) and 93.1% are present in the thinner one (factor 2) as shown in Fig. 13b. Criterion performances in the unaged state are consistent with the literature results for composites conditioned under less severe conditions (DAM or in range of $T_g$+[30°C, 45°C], [16,17]). Meaning that, for this criterion, performances do not depend on the test temperature. As shown by the figure, aging does not impact the criterion unifying capacities. The fatigue curve is simply shifted towards lower values with a slight decrease of slope, and a smaller intercept due to the gap between the curves. The gap for this FIP is greater than the one given with the cyclic mean strain rate-based criterion due to the difference of the mean stress in the energy computation. This was expected since stress amplitudes were dependent on the orientation and fiber content (Section 3.2.2).

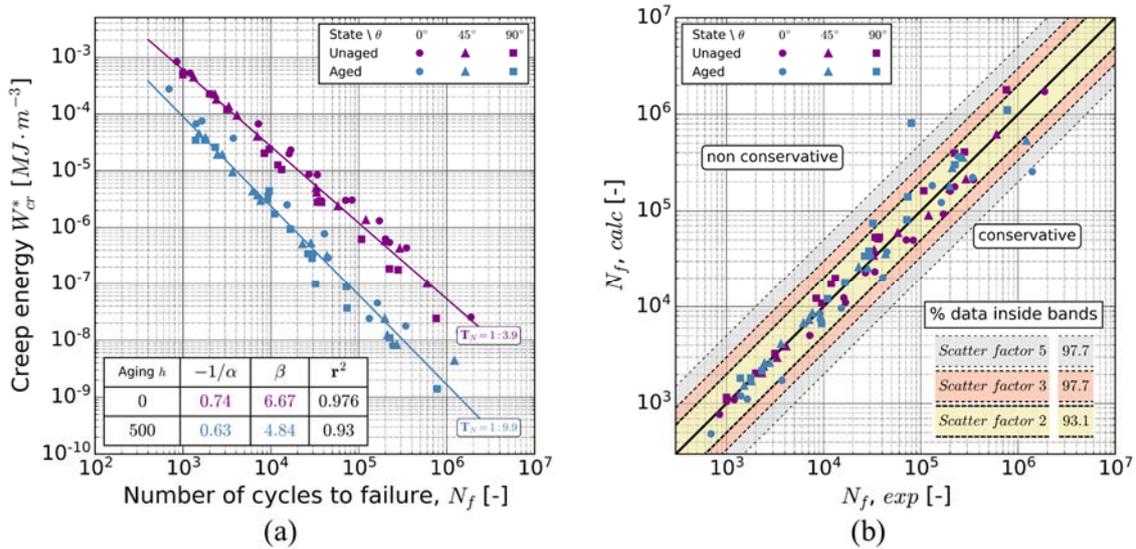

*Figure 13 Performance of the cyclic creep energy based criterion: (a) fatigue curve and identified parameters for each conditioning (unaged-aged) and (b) simulated number of cycles, Nf,calc, vs. experimental number of cycles to failure, Nf,exp - 35% wt. fiber composite.*

Results still obtained with the mid-life cyclic creep energy ($W_{cr}^*$) based criterion for the 50% wt. composite are shown in Fig. 14. Comparable performances are exhibited by this criterion with 97.5% and 92.5% of data points in the bands of factors 5 and 3, respectively, while a slight drop is observed in the thinner one (73.8%). These results are consistent with the previously observed differences between the two fiber contents (S-N curves, Fig. 7). With less packed data (both before and after aging), linked to an aging impact less noticeable at 0° and 45°, the overall dispersion appears larger in this case.

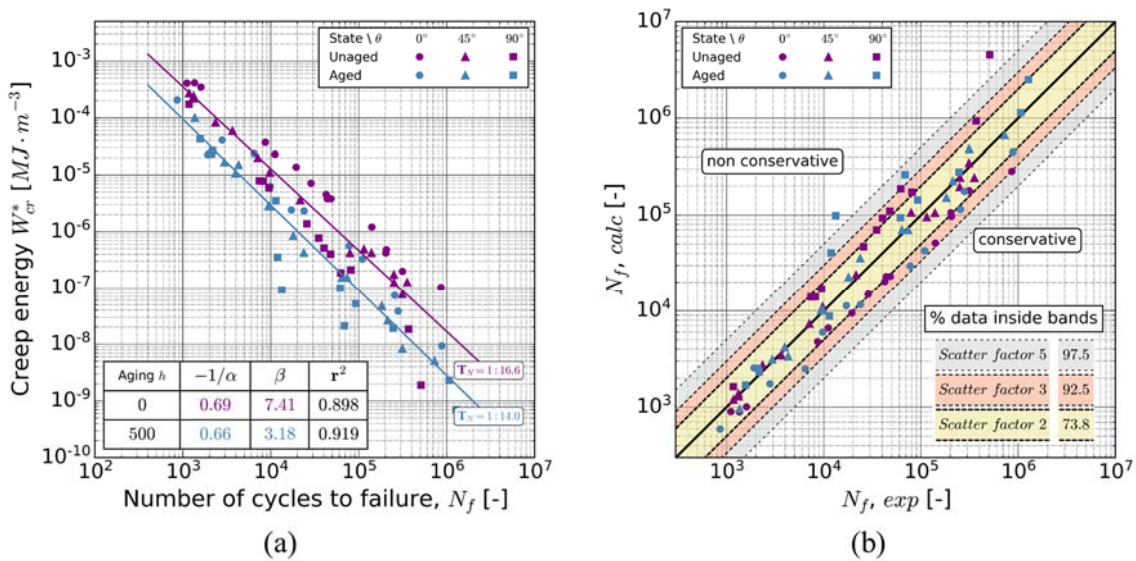

*Figure 14 Performance of the cyclic creep energy criterion: (a) fatigue curve and identified parameters for each conditioning (unaged-aged) and (b) simulated number of cycles, Nf,calc, vs. experimental number of cycles to failure, Nf,exp - 50% wt. fiber composite.*

Following the above-described strategy for the mid-life cyclic creep energy ($W_{cr}^*$) based criterion, percentages of data points found inside each scatter band (factor 2, 3, and 5) were reported horizontally in a histogram chart for every criterion (see Table 1). The results were plotted for both composites in Fig. 15. Percentages of data points in each scatter band are displayed for each conditioning separately (on both sides of the vertical middle axis). Percentages of the largest scatter band (factor 5) are written at the extremity of the bars. The results will be discussed first for the 35% wt. composite (Fig. 15a) and compared to those of the other material later on (Fig. 15b).

In the same way as for the unaged materials reported in the literature, the classic S-N criterion ($\sigma_{max}$) fails at unifying the data. No data points are found inside any band for the unaged material while around 33% are found in the largest band (factor 5) for the aged one as shown in Fig. 15a. Normalization by the ultimate tensile stress ($\sigma_{UTS}$) increase the accuracy of the criterion with 97.6% and 95.5% of data in the larger band for the unaged and aged material, respectively. However, only 54.8% and 68.8% are found in the thinner band (factor 2) for the unaged and aged material, respectively. The energetic-based criteria ranking order in the unaged state is found to be consistent with the literature results for this load ratio (i.e., $R_\sigma = 0.1$). The mid-life elastic strain energy criterion ($W_e^*$) is less good at data unification (factors 5 and 2, 61.9% and 35.8%) while better results are obtained with the mid-life hysteresis energy ($W_h^*$) (factors 5 and 2, 92.9% and 59.5%). Lastly, as shown previously (Fig. 13-14), the mid-life cyclic creep energy ($W_{cr}^*$) criterion exhibits the best results out of all power-law tested criteria. In the aged state, the same ranking order is found with slightly better results in most scatter bands. For instance, results obtained with $W_e^*$ and $W_h^*$ are at 75.5% and 93.3% in the band of factor 5, respectively (31.1% and 53.3% in the factor 2 band). The remaining criteria (mixed-quantity) needed an optimization method for parameters identification which was performed following the authors' recommendations, see [16,17]. Their performances fall in line with those obtained for the $W_{cr}^*$ based criterion, while having a higher cost-efficiency ratio, which is not beneficial for the small available database. The situation would be different for a larger database including several positive and negative load ratios.

Comparable results were obtained for the 50 % wt. material (Fig. 15b), with fewer data located in the scatter bands of factors 2 and 3 mostly due to a higher data dispersion. Data unifications for the aged material are slightly better for criteria based on the cyclic creep energy. This is linked to the data scattering between the three orientations, larger in the case of the unaged material as shown in Fig. 12b.

Overall, these results demonstrate that criteria performances are slightly dependent on the conditioning state even with irreversible changes. In the case of a fully aged material, in which degradation has reached a sufficiently advanced stage (i.e., 35% wt. composite), similar criteria performances as for unaged material are obtained as expected. However, intermediate stages such as observed in the 50% wt. composite should rather be tackled using criteria whose FIP evolution is less impacted by aging such as the cyclic creep energy.

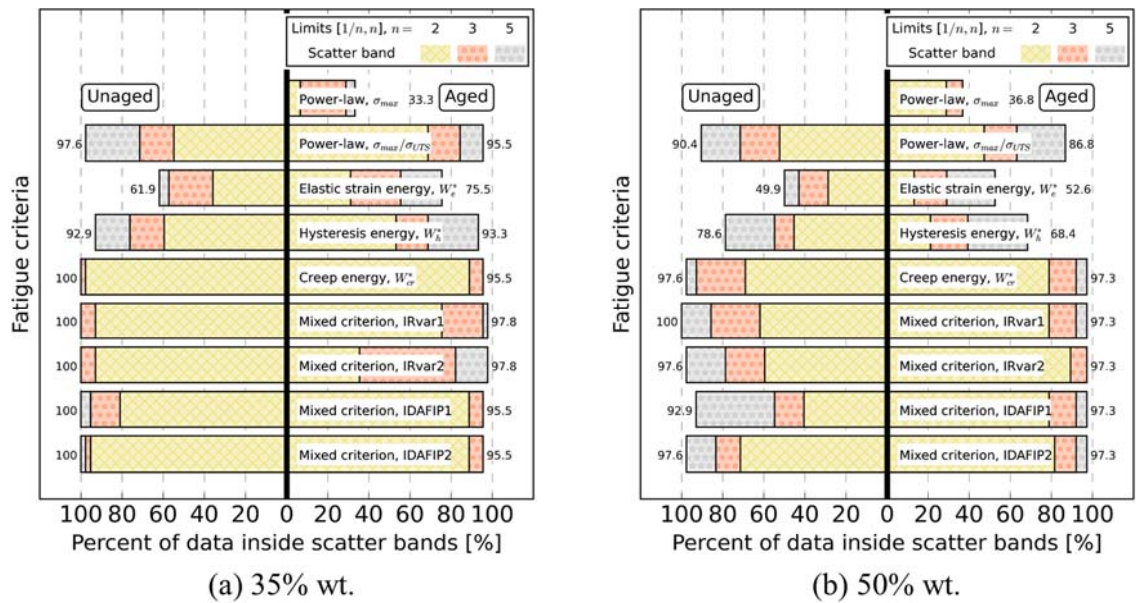

*Figure 15 Criteria unifying capacities: histogram representation showing the data presence in each scatter band (factors 2, 3, and 5). The parameters of each criterion were identified for each conditioning separately by using the 3 orientations: (a) 35% wt. fiber composite; (b) 50% wt. fiber composite.*

### 4.2 Prediction of fatigue lifetime of aged materials from criteria identified on fatigue data of unaged materials

In an industrial context, generating systematically a database for both unaged and aged materials under peculiar conditions is often time- and resource-consuming especially when developing new materials. Despite the existence of means of performing accelerated aging [31], such anisotropic materials require loadings along several orientations, which inevitably produce a large database. Hence, being able to predict fatigue lives of aged materials using only fatigue data in the unaged state seems attractive. The relevance of such a strategy is here explored. The criteria predicting capacities are appreciated using the computed number of cycles to failure ($N_{f,calc}$) against the experimental data ($N_{f,exp}$) and investigating the data dispersion within the scatter bands. The results, for the mid-life cyclic creep energy ($W_{cr}^*$) are presented in Fig. 16 for both fiber contents. As expected from using the parameters of the unaged material, the data points of the aged material appear shifted in the non-conservative domain as the unaged average lifetimes (derived from the regression lines) tend to surpass the aged ones for each and every criterion. The motivation here, is to see if the shift is the same in each case and consistent with the previous observations. Since curves were already overlapped for the 50% wt. composite, a good portion of data points in the aged state fall within the scatter band of factor 3 (75%) as shown in Fig. 16a. For the 35% wt. composite, only 52.9% of data are located in this very scattered band, essentially composed of unaged data points. The compiled performances of the selected fatigue criteria are reported in Fig. 17 similarly to Fig. 15.

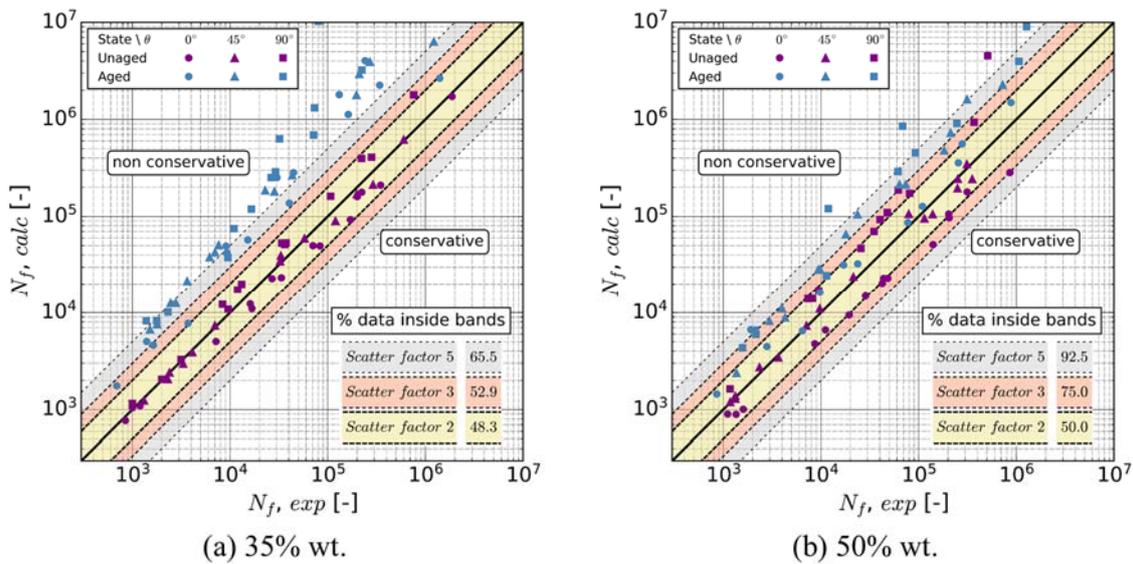

*Figure 16 Predicting capacities of the cyclic creep energy criterion: simulated number of cycles, $N_{f,calc}$, vs. experimental number of cycles to failure, $N_{f,exp}$.. The fatigue criterion parameters were identified on the unaged material database using the 3 orientations: (a) 35% wt. fiber composite; (b) 50% wt. fiber composite.*

For the aged 35% wt. material fatigue lives are poorly estimated by every tested criterion (Fig. 17a, left side). Out of the single-variable criteria, only 33.4% of data for the aged material are in the largest band with the cyclic creep energy criterion. The mixed criteria exhibit close performances related to this latter criterion. Better results are obtained for the 50% wt. material as shown in Fig. 17b. Almost 35% and 87% of data points for the aged material using the cyclic creep energy criterion are found in the scatter bands of factors 2 and 5, respectively. Other single-variable criteria fall with around 5% of predicted fatigue lives in the larger scatter band. This noticeable difference in data prediction, for the two composites, can be attributed to the data dispersion and curves clustering of the most aging resilient orientations of the 50% wt. material (0 & 45°).

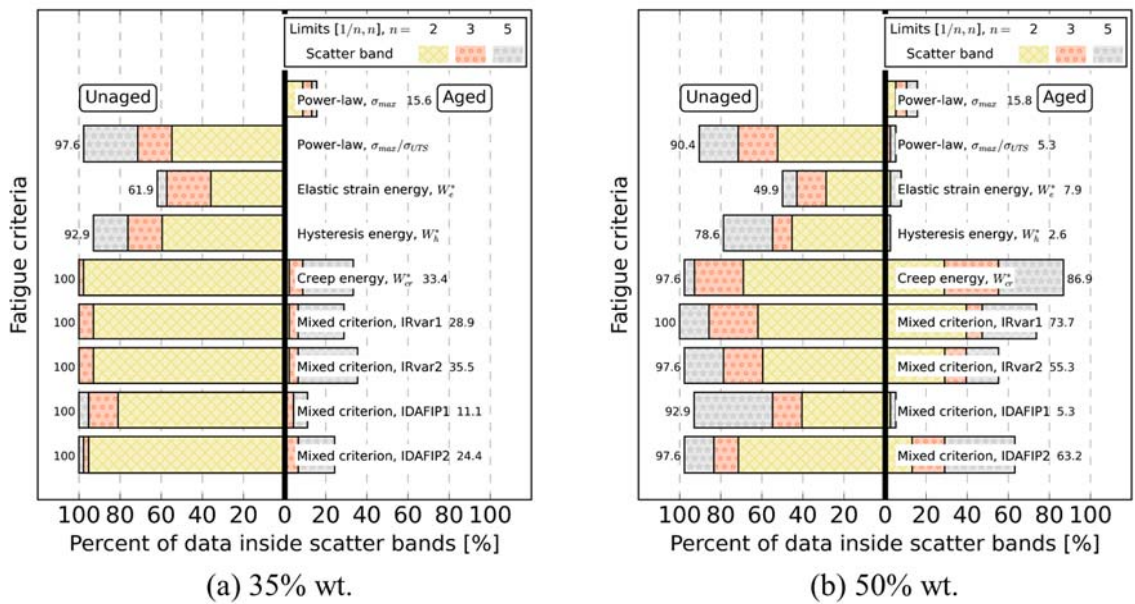

*Figure 17 Criteria predicting capacities: histogram representation showing the data presence in each scatter band (factors 2, 3, and 5). The parameters of each criterion were identified on the unaged material database using the 3 orientations: 35% wt. fiber composite; (b) 50% wt. fiber composite.*

## 5. Conclusions

The main objective was to study the severe thermo-oxidative aging impact on the tensile and fatigue response of polyamide-based materials (matrix and composite). Two composites with two fiber contents (35, 50% wt.) were compared to study the microstructure influence. Tests were performed on unaged and aged (200 °C, 500h) materials at aging temperature (200°C), for the bare matrix and all three orientations of the composite materials. Tensile tests were performed for both matrix and composite specimens to be compared with cyclic data. The fatigue database was built for one load ratio ($R = 0.1$) and exclusively for the composite materials. The main results obtained are the following:

- Monotonic tests in pure matrix specimens confirmed the stiffening and embrittlement effect of thermo-oxidative aging in polyamides. Comparison of matrix and composite specimens showed that the introduction of fibers as reinforcements helps to reduce the overall effect of aging on the strain and stress properties of the material even though the interface is also subjected to aging degradation. Fiber orientation also plays a part in the aging process, the softest materials, with fibers mostly perpendicular to the tensile direction, tend to be more degraded;
- The increase of fiber content (from 35% wt. to 50%) reduces the overall aging impact with partial preservation of mechanical properties (strength; strain at break) with increased fatigue lives at a given stress level. Specimens with fibers more aligned with the loading directions seemed to benefit the most from such an increase;
- Failure mechanisms were analyzed from SEM observations. Combined with the strain evolution, they revealed that the initiation approach for fatigue lifetime estimation is valid even for aged specimens;
- Several cyclic indicators (maximum stress, elastic strain energy, hysteresis energy, and creep energy) were monitored and accounted for in the computation of several fatigue criteria reported in the literature. The best criteria for unaged materials, those based on the cyclic creep energy, were shown to still be the best for aged materials. When identified for each database separately (unaged; aged), the criteria abilities at unifying the corresponding data are similar for both states;
- Identifying the fatigue criterion on the unaged material and applying it to the aged ones, leads to poor results independently of the chosen criteria and material's fiber content. Work is on progress to evaluate separately the improvement resulting from the introduction of the constitutive law of the aged matrix and/or a modification of the fatigue criterion.

Finally, other loading ratios need to be investigated, to confirm that, criteria performances remain consistent after thermo-oxidation. It would also be interesting to verify if for other aging conditions, particularly in the presence of water, the same observations can be made. Other specimen geometries (e.g., tubular) should be studied as well to generalize these conclusions in the case of multi-axial loading.

**Acknowledgments**

Pprime Institute gratefully acknowledges «Contrat de Plan Etat - Région Nouvelle-Aquitaine" (CPER) as well as the "Fonds Européen de Développement Régional (FEDER)" for their financial support to the reported work. The authors would like to thank the ANRT (French National Association for Research and Technology) for its financial support via a CIFRE grant.